\documentclass[sigconf]{acmart}

\usepackage{booktabs} % For formal tables
\usepackage{color, colortbl}

\setcopyright{rightsretained}

\begin{document}
\settopmatter{printacmref=false} % Removes acm reference below abstract
\settopmatter{printccs=false} % Adds  CCS indexing information below abstract
\renewcommand\footnotetextcopyrightpermission[1]{} % removes footnote with conference information in first column
\pagestyle{plain}

\title{Improved Answer Selection with Pre-Trained Word Embeddings}

\author{Rishav Chakravarti}
\affiliation{
  \institution{IBM Watson}
}
\email{rchakravarti@us.ibm.com}

\author{Ji\v{r}\'\i\/ Navr\'atil}
\affiliation{
  \institution{IBM Watson}
}
\email{jiri@us.ibm.com}

\author{C{\'i}cero Nogueira dos Santos}
\orcid{1234-5678-9012}
\affiliation{
  \institution{AI Foundations, IBM Research}
}
\email{cicerons@us.ibm.com}

\begin{abstract}
 This paper evaluates existing and newly proposed answer selection methods based on pre-trained word embeddings. Word embeddings are highly effective in various natural language
 processing tasks and their integration into traditional information retrieval (IR) systems
 allows for the capture of semantic relatedness between questions and answers. 
 Empirical results
 on three publicly available data sets show significant gains over traditional term frequency based
 approaches in both  supervised and unsupervised settings.  
 We show that combining these word embedding features with traditional learning-to-rank techniques can achieve similar performance to state-of-the-art neural networks trained for the answer selection task.
 
\end{abstract}




\maketitle

% Color declaration for table formatting
\definecolor{light-gray}{gray}{0.95}

\section{Introduction}
A core objective of Question Answering (QA) systems is to maximize the utility of selected candidate answers with respect to the user's question. Often, QA systems implement a solution in two phases:
\begin{enumerate}
\item an initial retrieval phase based on term overlap with the (expanded) question
\item a secondary ranking of the candidates to maximize relevance of the top-k answers with respect to the question
\end{enumerate}

Traditionally, secondary ranking uses language or relevance modeling approaches that still rely on term overlap statistics between the (expanded) question and the answer text \cite{Lavrenko:2001:RBL:383952.383972, Macdonald:2013:WHL:2559123.2559126}. Reliance on term overlap can suffer from lexical gaps between the language used to express the question and the language used in the candidate answer text even when there are semantic matches. 

Word embeddings such as Word2Vec \cite{mikolov2013distributed} and GloVe \cite{pennington2014glove} have surfaced as a way to model the semantics behind terms in various natural language processing tasks. We evaluate existing and newly proposed approaches that integrate such pre-trained word embeddings into the ranking phase of the QA pipeline. We show experimental results in both supervised and unsupervised settings with comparisons to traditional IR as well as a state-of-the-art deep learning system.

The main contributions of this work are: (1) a thorough comparison of recently proposed methods to integrate word embeddings in the QA pipeline; (2) the proposal of an extension to the Relaxed Word Mover's Distance method that incorporates matching terms proximity in the answer text; (3) an empirical demonstration that combining traditional features with word embedding based features can boost the result of learning to rank approaches.

Section \ref{sec:relatedWork} reviews the existing word embedding based approaches for ranking which we evaluate in Section \ref{sec:results}. In addition, Section \ref{sec:proposed} proposes additional techniques to address generalizability while maintaining performance. Sections \ref{sec:experimentalSetup} and \ref{sec:results} describe the experimental setup and results, respectively. Finally, in Section \ref{sec:conclusions} we summarize findings on combining traditional IR techniques with word embedding based features. 

\section{Related Work}\label{sec:relatedWork}

In \cite{Zuccon:2015:IEN:2838931.2838936}, the authors propose using cosine similarity between embeddings for two words as an estimate of \textit{translation probability} between them. The embeddings are pre-trained using a large text corpus using Word2Vec. This \textit{translation probability} is then plugged into a language modeling framework that ranks candidate answers based on the likelihood score of each candidate answer producing the question terms (with some smoothing and collection effects taken into account). Their approach shows some improvements over a traditional language modeling baseline.

Work such as \cite{Zamani:2016:EQL:2970398.2970405} similarly incorporate word embeddings into the language model for the query \textit{expansion} task. Our work here on the ranking or answer selection task can be applied on top of any query expansion techniques. There is certainly overlap in the motivations for both tasks suggesting that both can be considered together in future work. 

Rather than \textit{sum} over all translation probabilities, \cite{2016arXiv160801972K,Brokos2016UsingCO} use a \textit{minimum} translation probability which they refer to as Relaxed Word Mover's Distance between the question and the candidate answer ($RWMD_Q$)\footnote{\cite{2016arXiv160801972K,Brokos2016UsingCO} also explore a \textit{first pass} search algorithm using the cosine similarity between the answer text word embeddings' centroids and the question word embeddings' centroid. In our preliminary experiments, this centroid based first pass search did not appear to provide a boost over traditional term based retrieval approaches. So we focus on the \textit{second pass} re-rank portion of their work}. Though effective, the authors note that their method does not differentiate between candidate answers which contain all question terms, so their results are reported on various hand-tuned hybrid models combining the $RWMD_Q$ score and traditional features like $BM25$. 

The authors in \cite{Lidan:Hybrid} utilize a similar hybrid model, where the word embedding score attempts to capture how much the candidate answer text changes with respect to the maximum and minimum values along each word embedding dimension. The degree of change is quantified by a weighted sum of the cosine similarity between the maximum and minimum pooled vectors of the original answer text and the answer text concatenated with the original question. We include the core component of this 'Min-Max Pooling' score ($MMP_{0.7}$) from \cite{Lidan:Hybrid} in our investigations. 

Finally, the work in \cite{Santos2016AttentivePN} represents state-of-the-art answer re-ranking using a fully supervised model that uses word embeddings as input features for a two-way attention pooling deep neural network trained on labeled question answer pairs in the usual learning-to-rank setup. There is a lot of similar work employing various neural network architectures such as \cite{Severyn:2015:LRS:2766462.2767738, Cohen:2016:EEL:2970398.2970438}.

In addition, in Section \ref{sec:proposed} we propose a term proximity based extension of the $RWMD_Q$ technique. There are numerous efforts to incorporate term proximity into traditional term coverage and language model based scores such as \cite{Tu:2013:EPF:2505515.2507864}. However, these approaches are based on the strict matching of terms in the question with terms in the answer text. In our work, we look at a simple approach which accounts for the \textit{soft} matching in the continuous word embedding space between two terms. In Section \ref{sec:results}, we evaluate this approach against the existing word embedding based answer selection strategies on three publicly available data sets. 

\section{Proposed Method}\label{sec:proposed}
\subsection{Unsupervised}\label{sec:unsupervisedProposals}

We propose a modification of $RWMD_Q$ called $Spanning-RWMD_Q$ ($S\mbox{-}RWMD_Q$) to take advantage of proximity in matching terms and to address a drawback observed during experimentation with large answer texts. 
The original $RWMD_Q$ algorithm performs scans over all words in the answer text to find the best matches (i.e. closest cosine distance in the embedding space) \textit{separately for each term}. For large answer text, this introduces the possibility of finding a good term matches for $RWMD_Q$ in unrelated contexts. Here is an example question with excerpts from two candidate answers in the result set:
\begin{itemize}
\item [Question] \textit{``Where to start looking for health insurance?''}
\item [Ans 1] \textit{``This debate has been going on for the last century. In theory, how could anyone be opposed to everyone having \textbf{health insurance}?...the real question is can a society afford universal \textbf{health insurance} and thus universal health care in the sense that most Americans have become accustomed to getting it? When you \textbf{start looking} at the economics of health care this becomes much more problematic...we have more MRI scanners in Memphis, TN than in all of Canada, a nation often cited as an example of \textbf{where} universal \textbf{health insurance} is a success...''}
\item [Ans 2] \textit{``The best place to \textbf{start looking} for \textbf{health insurance} is the internet. With free rate quotes available it's...''}
\end{itemize}

While both candidate answers contain \textit{good} matches between question terms and terms in the answer text, the second answer contains those matches within a single sentence rather than terms spread out among unrelated sentences.
\footnote{In \cite{Lidan:Hybrid} answers are limited to their first $k$ tokens to presumably address length. This truncation works well when applied to $RWMD_Q$ on the data set from which the above example is pulled. However, it was found inconsistent in our experiments on other data sets.}
To capitalize on any signals that may arise from the proximity of \textit{good} matches, $S\mbox{-}RWMD_Q$ first splits the answer text into overlapping spans of words and calculates a $RWMD_Q$ score on each span.

The maximum across the spans is then used as the final rank score: 
\begin{equation}
\smash{\displaystyle\max_{span \in doc}RWMD_Q(question, words_{span})}
\end{equation}

We also considered other approaches for dealing with large answer texts by (1) computing $RWMD_Q$ only on the top-$k$ TF-IDF terms in the answer text or (2) performing dimensionality reduction on the word embeddings in the answer pool prior to computing $RWMD_Q$ scores. The results are omitted due to space constraints, but were found to be inferior to the methods listed in Section \ref{sec:results}. 

\subsection{Supervised}
In addition to the standalone $S\mbox{-}RWMD_Q$ score, we also evaluate a combination of embedding based scores in a traditional learning-to-rank setup where these scores are passed in as features to a model trained using labeled data. 

%RISHAV: cutting for space
%The motivation is to draw on the observations from \cite{Lidan:Hybrid, Brokos2016UsingCO, 2016arXiv160801972K} that these features provide complementary information to traditional term-frequency based measures.

\section{Experimental Setup}\label{sec:experimentalSetup}
\subsection{Data Sets}
Experiments and analyses are carried out on three different publicly available Question-Answering (QA) data sets: Insurance QA, Home Depot, and BioASQ Task 4b.

All three data sets provide question text along with relevant answer ids from their respective corpora (see Table \ref{tab:dataSets} for statistics). For the BioASQ data set, the corpus consists of titles and abstracts for each article taken from MEDLINE/PubMed citation records\footnote{https://www.nlm.nih.gov/databases/download/pubmed\_medline.html}. We concatenate the title and abstract to be consistent with the original $RWMD_Q$ experiments presented in \cite{Brokos2016UsingCO}. 

\begin{table}
 \caption{Question-Answering Data Sets}
 \label{tab:dataSets}
 \begin{tabular}{ccl}
 \toprule
 &Doc Corpus Size & Labeled Queries\\
 \midrule
 InsuranceQA\footnote{\url{https://github.com/shuzi/insuranceQA} \cite{InsuranceQA}} & 27,413 & Train: 12,889 Test: 2,000\\
 Home Depot\footnote{\url{https://www.kaggle.com/c/home-depot-product-search-relevance/data}} & 124,412 & Train: 7,795 Test: 1,000 \\ 
 BioASQ\footnote{\url{http://participants-area.bioasq.org/general\_information/Task4b} \cite{tsatsaronis2015overview}} & 14,939,692 & 1,307 (split into 5 CV-folds)\\
 \bottomrule
\end{tabular}
\end{table}

\subsection{First Pass Result Sets}
Solr's\footnote{https://lucene.apache.org/core/5\_4\_0/core/org/apache/lucene/search/similarities/ LMDirichletSimilarity.html} implementation of the language modeling Dirichlet smoothed similarity score from \cite{Zhai:2001:SSM:383952.384019} produces first-pass result sets of 400 candidate answers\footnote{Results appeared to be robust to variations in the number of candidates} for each question. Solr is configured with answers in a single text field and analyzed for stop word removal, lower casing, English possessive term removal, and Porter stemming. 

\subsection{Baselines}
\subsubsection{Unsupervised}
\label{subsubsec:unsupervisedBaseline}
In addition to baseline first pass mentioned above, we also compare against a competitive second pass ranker based on traditional relevance modeling framework. Due to space constraints, we do not discuss the details here, but the system is based on work from \cite{Lavrenko:2001:RBL:383952.383972, Paik:2014:FMW:2661829.2661957}.

\subsubsection{Supervised}
\label{subsec:l2rBaseFeatures}
\label{subsec:l2rModel}
In the supervised setting, the traditional learning-to-rank baseline consists of a jforests implementation of LambdaMART \cite{Ganji:2011:SIGIR, burges2010ranknet} trained on the following term coverage based features: Unigram, Bigram, Skipgram, SloppyBigram, Solr Similarity Score. The model is trained using a variety of hyperparameter settings and the scores are averaged for robustness and consistency. In addition, a question-level standardized version of the feature values are appended to the raw feature set. 

For an end-to-end neural network based learning-to-rank model, we use an attentive pooling convolutional neural network (AP-CNN) presented in \cite{Santos2016AttentivePN}. A reasonable set of hyperparameter settings are chosen from experiments with the validation set from the InsuranceQA data set and then applied uniformly to the other two data sets. It is possible that further tuning on those data sets would yield marginally higher performs. 

%RISHAV: cut
%However, we didn't wish to further split the training corpora for those data sets (training the model already requires at least one split in order to choose the best epoch from training).

\subsection{Word Embeddings}
For the BioASQ data set, we use the pre-trained word embeddings provided by the authors of \cite{2016arXiv160801972K}\footnote{\url{https://www.ncbi.nlm.nih.gov/CBBresearch/Wilbur/IRET/DATASET/}}. It was learned on the PubMED corpus using a traditional skipgram word2vec model. For the remaining two data sets, we use a skipgram word2vec model with a context window size of 9 learned on the Wikipedia and Yahoo Answers corpus to produce word embeddings with 400 dimensions\footnote{Additional experiments were run using the pre-trained Google News word embeddings showing similar results}.

%RISHAV: cut since we talk briefly of tokenization earlier?
%\subsection{Text Pre-Processing}
%Stop words are removed and text is tokenized prior to calculating scores. We use the default tokenization provided with InsuranceQA, and use the gensim python library\footnote{\url{http://radimrehurek.com/gensim/utils.html}} for tokenizing the other data sets.

\subsection{Evaluation Metrics}
For InsuranceQA and Home Depot we report evaluation metrics on the test sets provided with each corpus (not the blind set provided with Home Depot). For BioASQ task 4b, we report an average over 5-fold cross-validation. 
We present results using both NDCG (Normalized Discounted Cumulative Gain) truncated at 20 and Precision at 1. 

\section{Experimental Results}\label{sec:results}
\begin{table*}
 \caption{Performance of Rankers}
 \label{tab:rankerPerformance}
 \begin{tabular}{lcc|cc|cc}
 \toprule
 & \multicolumn{2}{c|}{InsuranceQA} & \multicolumn{2}{c|}{HomeDepot} & \multicolumn{2}{c}{BioASQ}\\
 \midrule
 & $NDCG@20$ & $P@1$ & $NDCG@20$ & $P@1$ & $NDCG@20$ & $P@1$ \\
 \midrule
 \rowcolor{light-gray}
 \multicolumn{7}{c}{Unsupervised Models} \\
 \midrule
 
 1. $LM$ & 0.236 & 0.136 & $0.274^{[3]}$ & $0.224^{[3]}$ & 0.347 & 0.385 \\
 2. $LM + RWMD_Q$ & $0.286^{[1]}$ & $0.180^{[1]}$ & $0.299^{[3]}$ & $0.237^{[3]}$ & $\mathbf{0.385^{[1,3,5,6]}}$ & $0.414^{[1,3]}$ \\
 3. $LM + MMP_{0.7}$ & $0.288^{[1]}$ & $0.182^{[1]}$ & 0.230 & 0.168 & $0.349^{[1]}$ & $0.387^{[1]}$ \\
 4. $LM + S\mbox{-}RWMD_Q$ & $0.300^{[1,2,3,5]}$ & $0.196^{[1,2,3]}$ & $0.308^{[1,2,3]}$ & $0.243^{[1,3]}$ & $0.382^{[1,3,5,6]}$ & $\mathbf{0.428^{[1,3]}}$ \\
 5. $RM$ & $0.287^{[1]}$ & $0.192^{[1,2]}$ & $0.308^{[1,2,3]}$ & $0.253^{[1,3]}$ & $0.363^{[1,3]}$ & $0.403^{[1,3]}$ \\
 6. $RM+S\mbox{-}RWMD_Q$ & $\mathbf{0.324^{[1,2,3,4,5]}}$ & $\mathbf{0.218^{[1,2,3,4,5]}}$ & $\mathbf{0.324^{[1,2,3,4,5]}}$ & $\mathbf{0.254^{[1,3]}}$ & $0.368^{[1,3,5]}$ & $0.412^{[1,3,5]}$ \\ 
 \midrule
 \rowcolor{light-gray}
 \multicolumn{7}{c}{Supervised Models} \\
 \midrule
  7. $Base\mbox{-}\lambda M$ & 0.314 & 0.222 & 0.318 & 0.268 & $0.355^{[9]}$ & $0.386^{[9]}$ \\
 8. $Embedding\mbox{-}\lambda M$ & $\mathbf{0.377^{[7]}}$ & $\mathbf{0.304^{[7,9]}}$ & $\mathbf{0.363^{[7]}}$ & $\mathbf{0.328^{[7]}}$ & $\mathbf{0.398^{[7,9]}}$ & $\mathbf{0.437^{[7,9]}}$ \\
 9. $AP\mbox{-}CNN$ & $0.376^{[7]}$ & $0.272^{[7]}$ & 0.299 & 0.281 & 0.336 & 0.366 \\
 \bottomrule 
\end{tabular}
\caption*{Numerical superscripts indicate statistically significantly greater performance at $\alpha$ of 0.05 using a one-tailed paired t-test}
\end{table*}

The first section of Table \ref{tab:rankerPerformance} presents test performance of the following unsupervised ranking algorithms: 
\begin{enumerate}
\item $LM$ - First pass search ordering using Solr's language modeling implementation. The performance of this base retrieval is substantially higher than the base retrieval performance reported by \cite{Brokos2016UsingCO} on the BioASQ data set (perhaps due to tunable parameters in base retrieval systems and/or differences in corpus pre-processing). This discrepancy, along with the fact that we report numbers on 5-fold cross-validated results, means that the numbers are not directly comparable to \cite{Brokos2016UsingCO} (though the gain due to $RWMD_Q$ seems consistent).
\item $LM+RWMD_Q$ - CombSUM of the min-max normalized first pass scores and Relaxed Word Mover's Distance 
%% JIRI: do we have doc to question? was here: (question to doc) 
\item $LM+MMP_{0.7}$ - CombSUM of the min-max normalized first pass scores and a weighted average of the min and max pooling scores as per \cite{Lidan:Hybrid}. 
The blending weight was set to 0.7 and only the first 20 tokens of the document were taken as suggested in \cite{Lidan:Hybrid}. 
Furthermore, we wish to point out that our investigation includes a reproduction of $MMP_{0.7}$ as described in \cite{Lidan:Hybrid}, however, the authors advised us 
that their experimental results include additional robustness steps, including ensembling \cite{Lidan:Personal042017}.
%Also, in discussion with the authors, there were found to be a number of robustness tweaks (e.g. ensembling a set of randomly down sampled set of embedding vectors, base system reciprocal rank weighting etc) which were applied but not discussed at length in \cite{Lidan:Hybrid}. These robustness tweaks are definitely a topic for investigation in future work given the boosts that can be seen going from the base $MMP_{0.7}$ score here to the tweaked version of the score for which numbers are reported in \cite{Lidan:Hybrid}.

\item $LM+S\mbox{-}RWMD_Q$ - CombSUM of the min-max normalized first pass scores and the spanning version of the Relaxed Word Mover's Distance as described in section \ref{sec:unsupervisedProposals}. Spans consist of up to 20 consecutive tokens with the starting point of each span moving forward 2 tokens from the start of the previous span.
\item $RM$ - A competitive relevance modeling ranker based on traditional techniques discussed in section \ref{subsubsec:unsupervisedBaseline}.
\item $RM+S\mbox{-}RWMD_Q$ - CombSUM of the min-max normalized scores from $RM$ and $S\mbox{-}RWMD_Q$
\end{enumerate}

%RISHAV: Cut this
%We also explored an alternative way of combination by performing a (stable) re-sort of the first pass rankings using the various embedding based scores. While this still provides an improvement on certain data sets, it appears to be less effective than CombSUM fusion of the first pass scores.

%particularly when combined with $RM$ which is a strong baseline in itself. 

$LM+S\mbox{-}RWMD_Q$ provides a statistically significant boost 
over the first pass results across all data sets and also provides a statistically significant boost over the previously proposed $RWMD_Q$ and $MMP_{0.7}$ embedding based hybrid scores. Either the $RM+S\mbox{-}RWMD_Q$ or the $LM+S\mbox{-}RWMD_Q$ scores provides the best unsupervised model in all three data sets. We also tried combining the $LM$ baseline with proximity based metrics such as $MinDist$ \cite{Tu:2013:EPF:2505515.2507864} and $Spanning\mbox{-}TF\mbox{-}IDF$ (where $TF\mbox{-}IDF$ scores are calculated on each span rather than the full answer text) to see if simply taking proximity into account alone can explain the boost in performance. The results are omitted due to space constraints, but the boosts in performance were inferior to combining proximity with $RWMD_Q$.

In the second section of Table \ref{tab:rankerPerformance} we present the test performance of the following supervised ranking algorithms:
\begin{enumerate}
\item[(7)] $Base\mbox{-}\lambda M$ - Traditional learning-to-rank model trained using the base IR features discussed in section \ref{subsec:l2rBaseFeatures} and the LambdaMART algorithm discussed in \ref{subsec:l2rModel}.
\item[(8)] $Embedding\mbox{-}\lambda M$ - Traditional learning-to-rank model trained using the base IR features discussed in section \ref{subsec:l2rBaseFeatures} plus the embedding based features from Table \ref{tab:rankerPerformance}: $RWMD_Q$, $MMP$\footnote{Rather than combine the min and max pool scores from \cite{Lidan:Hybrid} using the fixed weight, we keep the min and max pool scores separate}, and $S\mbox{-}RWMD_Q$. We use the LambdaMART algorithm discussed in \ref{subsec:l2rModel}.
\item[(9)] $AP\mbox{-}CNN$ - The attentive-pooling convolutional neural network discussed in \ref{subsec:l2rModel}.
\end{enumerate}

Incorporating the embedding based features into a LambdaMART model provides a statistically significant boost over just using traditional learning-to-rank features as well as the best-performing unsupervised technique. It even surpasses the $AP\mbox{-}CNN$ neural network model, though the difference only appears to be statistically significant in the third data set where the drastically smaller data set size may be hindering the neural network's capacity to learn.

\section{Conclusion}\label{sec:conclusions}
In this paper, we show that combining word embedding based re-ranking strategies with traditional IR techniques provides significant boosts in performance. Specifically, a standard supervised learning-to-rank model can combine word embedding based features with traditional term coverage based features to produce ranking accuracy that is comparable to state-of-the-art data- and training-intensive techniques like attentive pooling convolutional neural networks.

The proposed method provides a boost over traditional relevance modeling based approaches in the absence of labeled training data. In fact, this unsupervised approach appears to even outperform supervised approaches using term-coverage (though perhaps additional feature engineering could be carried out to boost the baseline used in this paper). The $S\mbox{-}RWMD_Q$ variant also consistently outperforms existing word embedding based features both in the unsupervised as well as the supervised settings.

% TODO: ok to mention the Dual Embedding Space Model from Microsoft like so?
In our future work, we plan to investigate approaches which look at question expansion and term weighting in the embedding space as well as additional sources of word embeddings such as the work presented in \cite{Nalisnick:2016:IDR:2872518.2889361}.

\bibliographystyle{ACM-Reference-Format}
\bibliography{references.bibtex} 

\end{document}